# Topology Optimization of Surface-enhanced Raman Scattering Substrates

Ying Pan[1,2+], Rasmus E. Christiansen[3,4+], Jerome Michon[1,2], Juejun Hu[1,2], Steven G. Johnson[3*]

[1]Department of Materials Science and Engineering, Massachusetts Institute of Technology, Cambridge, MA 02139, USA

[2]Materials Research Laboratory, Massachusetts Institute of Technology, Cambridge, MA 02139, USA

[3]Department of Mathematics, Massachusetts Institute of Technology, Cambridge, MA 02139, USA

[4]Department of Mechanical Engineering, Technical University of Denmark, Nils Koppels Allé, Building 404, 2800 Kongens Lyngby, Denmark

+These authors contributed equally to this work.
*To whom correspondence should be addressed.

ABSTRACT

Surface-enhanced Raman spectroscopy is a powerful and versatile sensing method with a detection limit down to the single molecule level. In this article, we demonstrate how topology optimization (TopOpt) can be used for designing surface enhanced Raman scattering (SERS) substrates adhering to realistic fabrication constraints. As an example, we experimentally demonstrated a SERS enhancement factor of $5 \times 10^4$ for the 604 cm$^{-1}$ Raman line of rhodamine 6G using metal nanostructures with a critical dimension of 20 nm. We then show that, by relaxing the fabrication constraints, TopOpt may be used to design SERS substrates with orders of magnitude larger enhancement factor. The results validate topology optimization as an effective method for engineering nanostructures with optimal performance and fabrication tolerance.

INTRODUCTION

We demonstrate the power of topology optimization (TopOpt)[1–3] as a tool for inverse design of direct-lithography manufacturable periodic surface enhanced Raman Scattering (SERS) metasurfaces[4–11] through design, fabraction, and experimental validation. We show that imposing fabrication limitations incurred when using electron-beam-lithography (EBL) directly as part of the TopOpt procedure (an out-of-plane sidewall slant angle and minimum-feature-size of design members) ensures a close correspondence between the design blueprint and the fabricated structure, without the need for post-lithography processing. By relaxing the minimum-feature-size constraint we demonstrate numerically that a novel nanostructure is designed exhibiting a two-orders-of-magnitude increased Raman enhancement, hereby showing that with improved fabrication techniques it may be possible to approach the theoretical limit of SERS[12], which has hitherto proven illusive. Conversely, imposing strict fabrication limitations results in an optimized design resembling the well-known bowtie-antenna design[13], and unsurprisingly we observe similar experimental SERS enhancements.

Surface-enhanced Raman scattering (SERS) capitalizes on the local electromagnetic field enhancement near plasmonic metal nanostructures [4–11] to dramatically boost Raman scattering from molecules, enabling detection sensitivity down to the single molecule level[8,11,14–16]. However, such gaint enhancement is only found in structures with extremely fine feature size not compatible with standard lithographic fabrication.[16–19] In this paper, we applied topology optimization with realistic fabrication constraints to engineer metal nanostructures and experimentally demonstrated SERS enhancement factor (EF) of $5 \times 10^4$ based on topology-optimized designs. While this EF is less than those reported in several previous works[18,20], it is achieved using direct lithographic patterning with a specified resolution limit of 20 nm and no post-lithography process is implemented to reduce the feature size. Our results therefore envisage a practical route for SERS substrate fabrication using standard lithographic fabrication tools to improve both processing yield and reproducibility.

Density-based topology optimization (TopOpt)[1,21] is a well-established numerical inverse-design method for creating highly-optimized freeform solutions to structural-design problems, applicable across a wide range of physical systems [2,22–25], where fabrication restrictions can be accounted for in the design process, e.g. by introducing minimum length-scales on the material phases [26,27]. When applying TopOpt, the physics is (most often) modelled using partial differential equations. The design problem is recast as a constrained continuous-optimization problem, which is solved efficiently using a gradient-based optimization algorithm, allowing for a vast design space with nearly unlimited design freedom[23]. Recently, interest in applying TopOpt for photonics and plasmonics has grown[28,29] with recent applications to optimize a nanophotonic demultiplexer[30], metasurface optics[31–34], plasmonic nano-antennas[35], plasmonic enhancement of thermal emission[36] and topological photonics[21] to name a few. The capability of TopOpt to handle a vast design space is critical to optimization of complex nanostructures, which is essential to approach the Raman enhancement limit[3,12,37]. While our prior work demonstrated how TopOpt can be used to design plasmonic nanoparticles with orders of magnitude greater enhancement than the standard bowtie antenna[3], that investigation focused on two dimensional (2-D) structures and thus practical fabrication limitations were not accounted for. Here, we extend our approach to complex 3-D structures with full consideration of fabrication-related non-idealities, such as limited feature sizes, non-vertical sidewall angles, and corner rounding.

First we design a periodic array of three-dimensional metallic nanostructures made of platinum (Pt), resting on a blanket Pt film in an air background. The goal of the design procedure is to tailor a structure in the periodic unit cell to maximize the emission enhancement from a Raman molecule at a fixed point in space and at a set of targeted wavelengths ($\lambda$). To achieve this goal, we employ density-based topology optimization as detailed in our previous work[3]. In brief, the design procedure assumes a fixed position of the Raman molecule, modelled as coupled dipole absorption and emission processes. We model the physics as a two-step process using Maxwell's equations for a time-harmonic electromagnetic field. First, the molecule is excited by an external source (at $\lambda = 532$ nm), followed by subsequent Raman emission of light from the molecule (at $\lambda = 549$ nm). The design domain consists of a brick-shaped spatial region with lateral dimensions 600 nm × 600 nm (equal to the lateral periodicity) and a height of 200 nm. which is discretized using a structured mesh. The lateral dimensions of the design domain were chosen to limit the effect of electron backscattering in the EBL process but was not otherwise optimized. The design is formed iteratively, by gradually changing the material distribution in each element of the mesh to either contain platinum or air, in order to maximize the Raman enhancement, while simultaneously respecting the constraints imposed on the model problem. In the present study we enforce an out-of-plane slant-angle on the design (incurred in the fabrication process) by employing a modified version of a filter and threshold procedure[26], as well as a minimum feature size of 20 nm (defined by the lithographic resolution) using a geometric length-scale constraint[27]. Details regarding the design process are found in the methods section.

The optimized nanostructure is depicted in Fig. 1. The black color indicates the Pt substrate, while the white color represents the optimized Pt nanostructures. Figure 1a presents a top-view of 3×3 unit cells for the optimized nanostructure and Fig. 1c shows a tilted view of the same structure. Even though we started the design process with a uniform Pt layer as an initial guess, the optimization led to a structure with a narrow gap in between a pair of bowtie-like nano-antennas. This optimized design geometry is consistent with the common notion that nano-gaps strongly localize the electromagnetic field[8–11], due to excitation of a localized plasmonic resonance combined with the corner singularity in the electric field near sharp tips[14–16], a phenomenon which appears instrumental to large Raman enhancement. Scanning electron microscopy (SEM) images in Fig. 1b and 1d illustrate the excellent fidelity of the nanostructures fabricated via a metal lift-off method, consistent with our design.

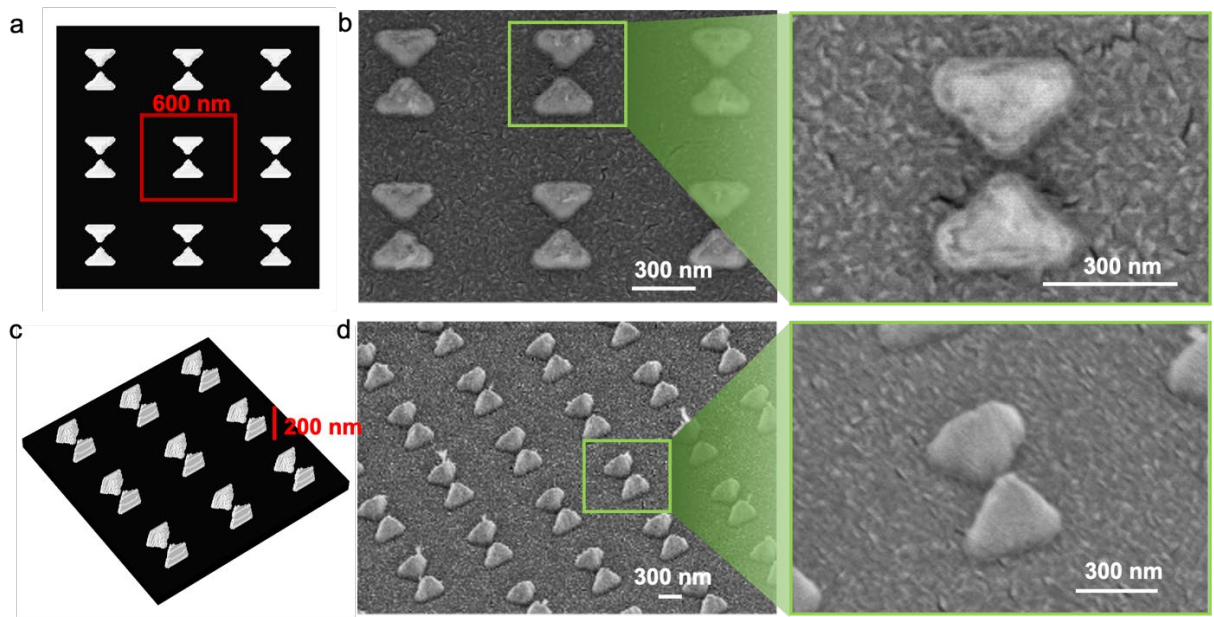

**Figure 1**. Images of the periodic array of the optimized nanostructures with the unit cell having a lateral dimensions of 600 nm × 600nm and a metal thickness of 200 nm. (a) Top view of the optimized nanostructures and (b) scanning electron microscopy (SEM) image of the fabricated nanostructures. (c) and (d) Tilted-view of the structures in (a) and (b). The insets on the right in (b) and (d) show zoomed in views of the structures. Regions with silver color represent the optimized
4

plastinum nanostructure and regions with black color indicate the platinum substrate in (a) and (c). The scale bar is 600 nm in (a), 200 nm in (c) and 300 nm in (b) and (d), respectively.

To experimentally investigate the Raman enhancement, Rhodamine 6G (R6G, Sigma-Aldrich) was chosen as the model molecule, whose Raman scattering response has been invesitgated in multiple prior studies[38–42]. R6G with a concentration of $10^{-3}$ M was prepared by dissolving 47.5 mg of R6G powder into 100 mL of deionized (DI) water. Aqueous R6G sulution with other concentrations are obtained by serial dilution of the initial solution in DI water. To transfer the R6G molecules onto the substrate, the platinum substrates were immersed in R6G solutions of varying concentrations for one hour and subsequently dried with nitrogen gas.

Raman spectra of R6G molecules on the substrates were measured using a Raman microscope (LabRam HR, HORIBA). Linearly polarized laser light at a wavelength of 532 nm was focused onto the substrate surface using a 50X objective lens. Backscattered light was collected by the same objective lens and directed to a spectrometer with 1800-g/mm grating. The laser spot size was approximately 1 μm, while the area containing the fabricated nano-structure array was 30 μm × 30 μm. The total power incident on the sample surface was 37±1 mW, measured at the sample holder stage. Data were collected at multiple locations for statistical averaging, with an accumulation time of 1 s for all measurements.

Figures 2a-b show the Raman spectra of R6G measured on the optimized SERS substrate and a bare Si substrate as a reference. The R6G was incubated at concentrations of $10^{-5}$ M (SERS) and $10^{-2}$ M (Si), respectively. The Raman spectrum in Fig. 2a features intense peaks at 604, 761, 1172, 1303, 1356, and 1647 cm$^{-1}$, which correspond to main vibrational features of carbon skeleton stretching modes in the R6G molecule[38,39,41,42]. The strongest band is seen at 604 cm$^{-1}$, whose intensity was targeted for maximization in the design process. This band is assigned to an in-plane bending mode of the xanthene ring[43].

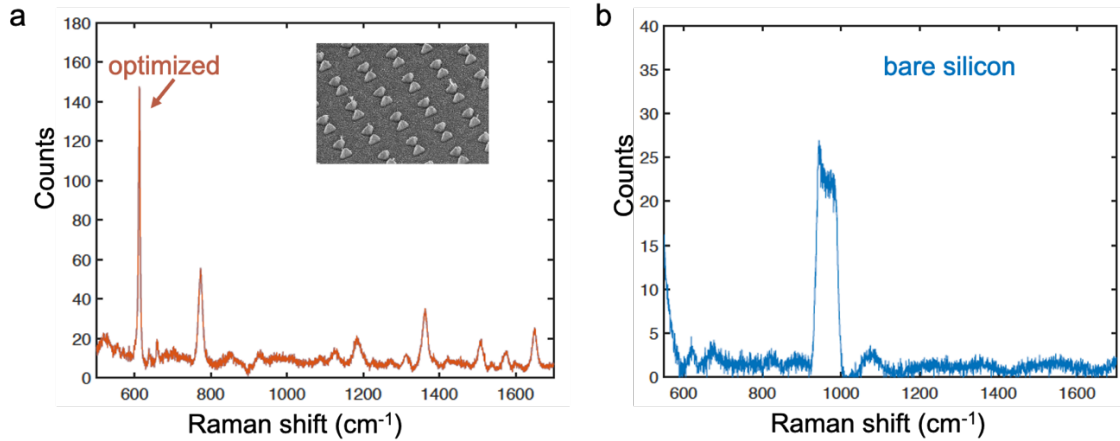

**Figure 2**. Comparison of Raman-spectra of R6G molecules on topology optimized (a) Pt nanostructures and (b) silicon substrate. The background was subtracted.

To evaluate the experimental SERS enhancement factor (EF), we compare the Raman signal measured on the nano-structured substrates with that from the reference silicon substrate. The EF is determined using the following equation:

$$EF = \frac{I_{SERS}}{C_{SERS}} \times \frac{C_0}{I_0}$$

where $I_{SERS}$, $I_0$ are the intensities of the 604 cm$^{-1}$ Raman peak measured on the SERS substrate and reference, respectively, and $C_{SERS}$ and $C_0$ are the concentrations of R6G in the aqueous solution that the SERS and reference substrates were treated with[14,42]. Using the formula, we obtained an average SERS EF of $5\pm0.3 \times 10^4$ for the 604 cm$^{-1}$ Raman peak. The measured EF is higher than that the EF-value of 4,300 predicted in our numerical simulations. The mismatch between the experiment and simulation is likely due to chemical enhancement[40,44], which provides additional Raman signal amplification due to electron transfer between R6G and the metal substrate. The magnitude of the chemical enhancement effect is in good agreement with previous reports, which point to one to two orders of magnitude of Raman signal enhancement through the chemical mechanism[45,46].

To demonstrate that the optimized nanostructure is robust against shape deviations and simultaneously compare the results to a well known SERS geometry, we fabricated a parametrized reference bowtie structure (Figs. 3d-e), whose shape follows the contour of the topology optimized structure (Figs. 3a-b), as shown in Figure 3. Experimental results in Figs. 3c and 3f suggest that the SERS EF does not vary significantly between the two realizations, when accounting for the measurement errors (below 15%).

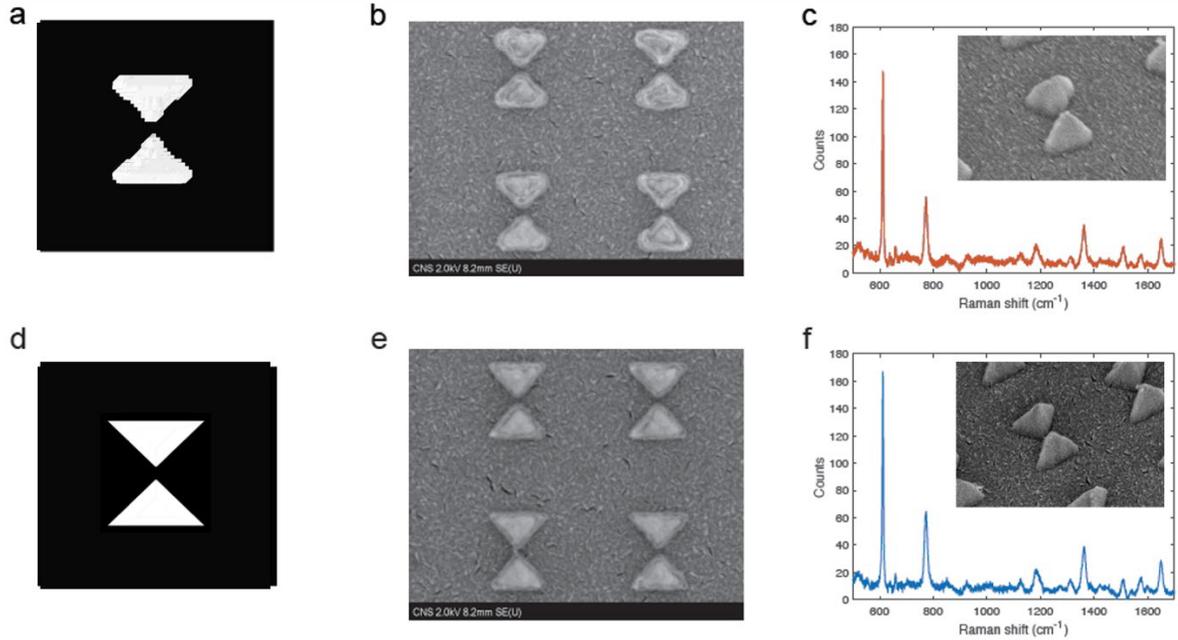

**Figure 3.** Nanostructure geometries and SERS EF measurement results for the topology optimized structure and a parametrised triangular bowtie nanostructure. (a) The optimized nanostructure unit cell, (b) SEM of the fabricated nanostructure array and (c) the measured Raman signal from the optimized nanostructure array in (b). (d)-(f) The corresponding information for the triangular bowtie nanostructures.

Finally, we demonstrate numerically that TopOpt can be used to design nanostructures with significantly higher Raman enhancement than that achieved by the nanostructure studied in Figs. 1-3. To this end, we relax the minimum-feature-size constarint from 20 nm to 5 nm. Furthermore, we reduce the lateral unit-cell area to 200 nm × 200 nm and increase the unit-cell height to 300 nm, hereby increasing the density of the Raman enhancement points by a factor of 9, as well as increasing the design freedom in the out-of-plane direction. Making these adjustmentsm we design the Pt nanostructure shown in Fig. 4. Figure 4a-c presents a tilted view, a top-down view, and a cross-secitonal view, respectively. The footprint of the designable unit cell is highlighted in Fig. 4b using a red box and the position of the Raman molecule is denoted by a red dot. The optimized nanostructure resembles an amalgamation of an in-plane bowtie-like antenna and an out-of-plane horn-like structure, suggesting that a combination of these two feature types may be exploited to increase the EF for other SERS structures. Figure 4d shows the nanostructure with a saturated colorscheme presenting the magnitude of the emitted power, shown in two vertical planes, for an array of Raman molecules placed at the point targed for Raman enhancement maximization. For the design in Fig. 4, the numerically calculated enhancement factor relative to the Raman molecule placed on a smooth silicon surface is 531,000, which (compared to the EF of 4,300 for the previous design) is an increase of roughly two orders of magnitude.

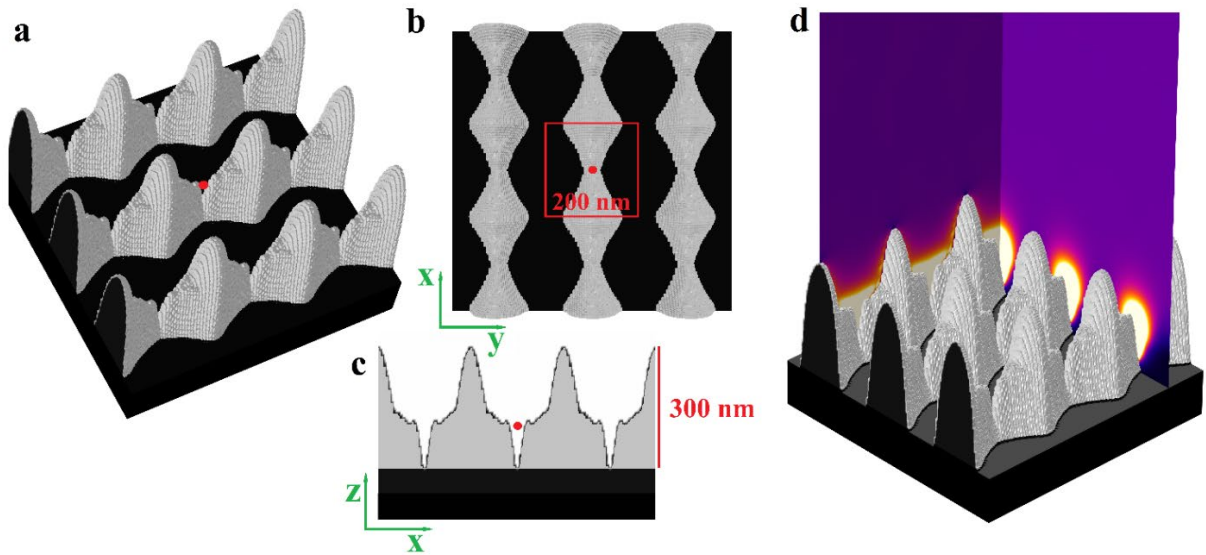

**Figure 4.** SERS nanostructure designed using TopOpt with relaxed fabrication limitations. (a) Tilted-view of the optimized nanostructure. (b) Top-down view of the nanostructure. (c) (x,z)-plane cross-sectional view of the nanostructure containing the position of the Raman molecule. (d) The magnitude of the powerflow around the nanostructures, shown in two vertical planes, for Raman molecules placed at the point(s) targeted for Raman enhancement maximization.

CONCLUSIONS AND OUTLOOK

We have demonstrated the experimental implementation of nanostructure arrays designed via topological optimization to maximize surface-enhanced Raman scattering. Our optimization approach incorporates approximations of practical fabrication constraints to generate realistic designs commensurate with direct lithographic patterning, which yields high EF values as validated in our experiment. These results illustrate the practical applicability and potential of applying TopOpt to Raman and other scattering problems to produce designs suitable for scalable manufacturing. Further, we demonstrated that TopOpt may be used to design nanostructures exhibiting even higher EFs by relaxing the fabrication constraints, motivating a hunt for improved fabrication techniques.

While traditional design techniques based on analytical models and intuition are able to solve relatively simple problems, such as designing SERS metasurfaces (e.g. by tailoring a bowtie-antenna array), our TopOpt-based method allows for envisioning and solving significantly more challenging design problems. For example, one can design of nanostructures tailored for maximizing Raman enhancement in a geometrically complex in-plane device, where analytical models and standard intuition cannot easily address the simultaneous objectives of emission enhancement and waveguide coupling[3,47]. More generally density-based topology optimization method offers a systematic approach for identifying high performance solutions to challenging design problems that balance multiple physical processes, enabling the creation of novel, complex devices.

## METHODS

### Inverse design of SERS substrate using Topology Optimization

The design domain for the topology-optimized nanostructure array in Figs. 1-3 is a unit cell with a lateral dimension of 600 nm × 600 nm and a height of 200 nm discretized using a uniform mesh with an element side length of 5 nm. For the array in Fig. 4 the unit cell has lateral dimensions of 200 nm × 200 nm and a height of 300 nm with an element side length of 2 nm.

The electromagnetic simulation is performed in COMSOL Multiphysics[48] using first-order finite elements. The optimization problem is solved using the Globally Convergent Method of Moving Asymptotes (GCMMA)[49]. For the physics modeling, the Raman molecule is modeled as a point dipole absorber/emitter at the center gap location. The wavelength of the incident field is 532 nm, while the wavelength of the emitted field is 549 nm. The emitted power, due to Raman scattering from the molecule, is calculated using an array-scanning method[3], which allows us to model incoherent dipole emission using a set of computations with Bloch-periodic boundary conditions. The relative permittivity of Pt is obtained from ellipsometry data measured on a J.A. Woollam M-88 multi-wavelength ellipsometer.

In the optimization process the Pt material distribution in the design domain is controlled using a two dimensional design field, $\xi(x,y)$, determining the cross-section of the Pt material layout. The design field is subjected to a smoothing scheme[26] using a standard cone-shaped filter and a filter radius of $r_f = 40$ nm, resulting in the filtered field $\tilde{\xi}(x,y)$. Following the smoothing operation, the design field is subjected to a height-dependent thresholding operation,

$$T(\tilde{\xi}(x,y),\eta(z),\beta) = \frac{\tanh(\beta \cdot \eta(z)) + \tanh\left(\beta \cdot \left(\tilde{\xi}(x,y) - \eta(z)\right)\right)}{\tanh(\beta \cdot \eta(z)) + \tanh(\beta \cdot (1 - \eta(z)))},$$

With a thresholding sharpness $\beta$ and thresholding level $\eta(z)$. This operation creates an extruded three-dimensional Pt-material layout (approximately) respecting the desired out-of-plane minimum side-wall slant-angle. This angle is specified via an interplay between the filter radius $r_f$ and the heigh-dependent thresholding level,

$$\eta(z) = \eta_1 + \left(\frac{(z - z_1)}{(z_2 - z_1)}\right) \cdot (\eta_2 - \eta_1).$$

Here $\eta_1 < \eta_2 \in [0,1]$ and $z_1(z_2)$ denotes the z-coordinate at the bottom(top) of the design domain.

In order to allow design freedom at the start of the optimization process and ensure a final (near) binary material distribution, the thresholding strength is increased gradually in the optimization process using the values, $\beta = \{8, 16, 32\}$.

As a final step, the filtered and thresholded design field is introduced into the model of the physics by using it to interpolate the relative permittivity in the design domain[50].

After the optimization process is complete the optimized permittivity distribution is sampled at the base of the design domain in (x,y)-coordinates with 0.5 nm resolution. The sampled field is smoothed with a 1.5 nm filter radius to eliminate kinks and the 50%-level contour extracted and used as the blueprint for fabrication.

Nanofabrication of the substrate
Platinum is initially deposited on the top of a silicon substrate using an electron beam evaporation system at a rate of 0.1 nm/s. The thickness of this Pt layer is 200 nm, which is sufficient to prevent interaction of the light with the underlying Si substrate. Then a layer of ZEP520A resist was spin coated (500 rpm, 5 s; 3000 rpm, 60 s) onto the substrate. Electron beam lithography (Elionix F-125) was used to define the structure. After exposure, the sample was developed in O-xylene for 2 min. The resulting pattern served as lift-off mask during subsequent metallic platinum evaporation step. A thin layer of titanium/platinum was deposited by electron beam evaporation and the thickness (10/200 nm) were confirmed by profilometer. The sample was immersed in acetone overnight to remove the resist, thereby forming platinum metallic structures on the substrate. The samples were cleaned by extensive rinsing in acetone and isopropyl before nitrogen gun drying.

## Acknowledgements

This work was financially supported by Danmarks Grundforskningsfond (DNRF147) through NanoPhoton – The Center for Nanophotonics, by the National Science Foundation under Award Number 1709212, and by the U.S. Army Research Office under award W911NF-18-2-0048. The sample fabrication was carried out at the Center for Nanoscale Systems (CNS) at Harvard University.

## Data Availability Statements

The data that supports the findings of this study are available within the article.

## Reference:

1. Bendsøe, M. P. & Sigmund, O. *Topology optimization: Theory, methods and applications*. (Springer, 2003).

2. Alexandersen, J. & Andreasen, C. S. A Review of Topology Optimisation for Fluid-Based Problems. *Fluids* **5**, 29 (2020).

3. Christiansen, R. E., Michon, J., Benzaouia, M., Sigmund, O. & Johnson, S. G. Inverse design of nanoparticles for enhanced Raman scattering. *Optics Express* **28**, 4444 (2020).

4. Mrozek, I. & Otto, A. *Surface enhanced Raman scattering*. *J. Phys.: Condens. Matter* vol. 4 (1992).

5. Kneipp, K. *et al. Single Molecule Detection Using Surface-Enhanced Raman Scattering (SERS)*. (1997).

6. Nie, S. & Emory, S. R. Probing single molecules and single nanoparticles by surface-enhanced Raman scattering. *Science* **275**, 1102–1106 (1997).

7. Kneipp, K., Kneipp, H., Itzkan, I., Dasari, R. R. & Feld, M. S. Surface-enhanced Raman scattering and biophysics. *Journal of Physics: Condensed Matter* **14**, 202 (2002).

8. Ru, E. C. L., Blackie, E., Meyer, M. & Etchegoint, P. G. Surface enhanced raman scattering enhancement factors: A comprehensive study. *Journal of Physical Chemistry C* **111**, 13794–13803 (2007).

9. Banholzer, M. J., Millstone, J. E., Qin, L. & Mirkin, C. A. Rationally designed nanostructures for surface-enhanced Raman spectroscopy. *Chemical Society Reviews* **37**, 885–897 (2008).

10. Ru, E. C. L. & Etchegoin, P. G. *Principles of surface-enhanced Raman spectroscopy : and related plasmonic effects*. (Elsevier, 2009).

11. Zhu, W., Banaee, M. G., Wang, D., Chu, Y. & Crozier, K. B. Lithographically fabricated optical antennas with gaps well below 10 nm. *Small* **7**, 1761–1766 (2011).

12. Michon, J., Benzaouia, M., Yao, W., Miller, O. D. & Johnson, S. G. Limits to surface-enhanced Raman scattering near arbitrary-shape scatterers. *Optics Express* **27**, 35189 (2019).

13. Hatab, N. A. *et al.* Free-standing optical gold bowtie nanoantenna with variable gap size for enhanced Raman spectroscopy. *Nano Letters* **10**, 4952–4955 (2010).

14. Li, K. *et al.* Free-standing Ag triangle arrays a configurable vertical gap for surface enhanced Raman spectroscopy. *Nanotechnology* **28**, (2017).

15. Caldwell, J. D. *et al.* Plasmonic nanopillar arrays for large-area, high-enhancement surface-enhanced Raman scattering sensors. *ACS Nano* **5**, 4046–4055 (2011).

16. Gong, T. *et al.* Highly reproducible and stable surface-enhanced Raman scattering substrates of graphene-Ag nanohole arrays fabricated by sub-diffraction plasmonic lithography. *OSA Continuum* **2**, 582 (2019).

17. Surbhi Lal, Stephan Link & Naomi J. Halas. Nano-optics from sensing to waveguiding. (2007) doi:10.1038/nphoton.2007.223.

18. Ward, D. R. *et al.* Electromigrated nanoscale gaps for surface-enhanced Raman spectroscopy. *Nano Letters* **7**, 1396–1400 (2007).

19. Gruber, C. M. *et al.* Fabrication of bow-tie antennas with mechanically tunable gap sizes below 5 nm for single-molecule emission and Raman scattering. *IEEE-NANO 2015 - 15th International Conference on Nanotechnology* 20–24 (2015) doi:10.1109/NANO.2015.7388978.

20. Gopinath, A. *et al.* Plasmonic nanogalaxies: Multiscale aperiodic arrays for surface-enhanced Raman sensing. *Nano Letters* **9**, 3922–3929 (2009).

21.  Christiansen, R. E. & Sigmund, O. A Tutorial for Inverse Design in Photonics by Topology Optimization. (2020).

22.  Christiansen, R. E. & Fernandez-Grande, E. Design of passive directional acoustic devices using Topology Optimization - from method to experimental validation. *The Journal of the Acoustical Society of America* **140**, 3862–3873 (2016).

23.  Aage, N., Andreassen, E., Lazarov, B. S. & Sigmund, O. Giga-voxel computational morphogenesis for structural design. *Nature* **550**, 84–86 (2017).

24.  Christiansen, R. E., Wang, F. & Sigmund, O. Topological Insulators by Topology Optimization. *Physical Review Letters* **122**, (2019).

25.  Lundgaard, C., Engelbrecht, K. & Sigmund, O. A density-based topology optimization methodology for thermal energy storage systems. *Structural and Multidisciplinary Optimization* **60**, 2189–2204 (2019).

26.  Wang, F., Lazarov, B. S. & Sigmund, O. On projection methods, convergence and robust formulations in topology optimization. *Structural and Multidisciplinary Optimization* **43**, 767–784 (2011).

27.  Zhou, M., Lazarov, B. S., Wang, F. & Sigmund, O. Minimum length scale in topology optimization by geometric constraints. *Computer Methods in Applied Mechanics and Engineering* **293**, 266–282 (2015).

28.  Jensen, J. S. & Sigmund, O. Topology optimization for nano-photonics. *Laser and Photonics Reviews* vol. 5 308–321 (2011).

29.  Molesky, S. *et al.* Inverse design in nanophotonics. *Nature Photonics* **12**, 659–670 (2018).

30.  Piggott, A. Y. *et al.* Inverse design and demonstration of a compact and broadband on-chip wavelength demultiplexer. *Nature Photonics* **9**, 374–377 (2015).

31.  Wang, E. W., Sell, D., Phan, T. & Fan, J. A. Robust design of topology-optimized metasurfaces. *Optical Materials Express* **9**, 469 (2019).

32.  Campbell, S. D. *et al.* Review of numerical optimization techniques for meta-device design [Invited]. *Optical Materials Express* **9**, 1842 (2019).

33.  Shalaginov, M. Y. *et al.* Design for quality: reconfigurable flat optics based on active metasurfaces. *Nanophotonics* **9**, 3505–3534 (2020).

34.  Chung, H. & Miller, O. D. High-NA achromatic metalenses by inverse design. *Optics Express* **28**, 6945 (2020).

35.  Wadbro, E. & Engström, C. Topology and shape optimization of plasmonic nano-antennas. *Computer Methods in Applied Mechanics and Engineering* **293**, 155–169 (2015).

36.     Kudyshev, Z. A., Kildishev, A. v., Shalaev, V. M. & Boltasseva, A. Machine-learning-assisted metasurface design for high-efficiency thermal emitter optimization. *Applied Physics Reviews* **7**, 021407 (2020).

37.     Miller, O. D. *et al.* Fundamental limits to optical response in absorptive systems. *Optics Express* **24**, 3329 (2016).

38.     Hildebrandt, P. & Stockhurger, M. Surface-Enhanced Resonance Raman Spectroscopy of Rhodamine 6G adsorbed on colloidal silver. *Journal of Physical Chemistry* **88**, 5935–5944 (1984).

39.     Shim, S., Stuart, C. M. & Mathies, R. A. Resonance Raman cross-sections and vibronic analysis of rhodamine 6G from broadband stimulated raman spectroscopy. *ChemPhysChem* **9**, 697–699 (2008).

40.     Michaels, A. M., Nirmal, M. & Brus, L. E. Surface enhanced Raman spectroscopy of individual rhodamine 6G molecules on large Ag nanocrystals. *Journal of the American Chemical Society* **121**, 9932–9939 (1999).

41.     Kudelski, A. Raman studies of rhodamine 6G and crystal violet sub-monolayers on electrochemically roughened silver substrates: Do dye molecules adsorb preferentially on highly SERS-active sites? *Chemical Physics Letters* **414**, 271–275 (2005).

42.     Zhang, P. *et al.* Large-scale uniform Au nanodisk arrays fabricated via x-ray interference lithography for reproducible and sensitive SERS substrate. *Nanotechnology* **25**, (2014).

43.     Watanabe, H., Hayazawa, N., Inouye, Y. & Kawata, S. DFT Vibrational Calculations of Rhodamine 6G Adsorbed on Silver: Analysis of Tip-Enhanced Raman Spectroscopy. *The Journal of Physical Chemistry B* **109**, 5012–5020 (2005).

44.     Wang, X., Huang, S. C., Hu, S., Yan, S. & Ren, B. Fundamental understanding and applications of plasmon-enhanced Raman spectroscopy. *Nature Reviews Physics* **2**, 253–271 (2020).

45.     H. Reilly, T., Chang, S.-H., D. Corbman, J., C. Schatz, G. & L. Rowlen, K. Quantitative Evaluation of Plasmon Enhanced Raman Scattering from Nanoaperture Arrays. *The Journal of Physical Chemistry C* **111**, 1689–1694 (2006).

46.     Banaee, M. G. & Crozier, K. B. Gold nanorings as substrates for surface-enhanced Raman scattering. *Optics Letters* **35**, 760 (2010).

47.     Kita, D. M., Michon, J. & Hu, J. A packaged, fiber-coupled waveguide-enhanced Raman spectroscopic sensor. *Optics Express* **28**, 14963 (2020).

48.     COMSOL Multiphysics v. 5.5,.

49.     Svanberg, K. A Class of Globally Convergent Optimization Methods Based on Conservative Convex Separable Approximations. *SIAM Journal on Optimization* **12**, 555–573 (2002).

50. Christiansen, R. E., Vester-Petersen, J., Madsen, S. P. & Sigmund, O. A non-linear material interpolation for design of metallic nano-particles using topology optimization. *Computer Methods in Applied Mechanics and Engineering* **343**, 23–39 (2019).